\documentclass[12pt,preprint]{aastex}

\usepackage{times}
\usepackage{lscape}

\newcommand{\ovi}{O$\;${\small\rm VI}\relax}

\newcommand{\sii}{S$\;${\small\rm II}\relax}
\newcommand{\feii}{Fe$\;${\small\rm II}\relax}

\newcommand{\fethree}{Fe$\;${\small\rm III}\relax}
\newcommand{\sthree}{S$\;${\small\rm III}\relax}

\newcommand{\lyalpha}{Lyman-$\alpha$}

\newcommand{\electron}{$e^-$}

\newcommand{\htwo}{H$_2$}
\newcommand{\HI}{H$\;${\small\rm I}\relax}
\newcommand{\HII}{H$\;${\small\rm II}\relax}

\newcommand{\Nelectron}{\ensuremath{N(e^-)}}

\newcommand{\kms}{km~s$^{-1}$\relax}

\newcommand{\mvlsr}{v_{\rm LSR}\relax}
\newcommand{\percc}{cm$^{-3}$\relax}
\newcommand{\column}{cm$^{-2}$}

\newcommand{\e}[1]{10^{#1}}

\newcommand{\abun}[1]{\ensuremath{A({\rm #1})}}

\newcommand{\fuse}{{\em FUSE}}

\newcommand{\hst}{{\em HST}}

\newcommand{\vz}{vZ~1128}

\newcommand{\subwim}{\ensuremath{_{WIM}}}
\newcommand{\subhim}{\ensuremath{_{HIM}}}

\newcommand{\pone}{Paper~I}

\begin{document}

\slugcomment{\em Accepted for publication in the Astrophysical Journal}



\title{A Method for Deriving Accurate Gas-Phase Abundances for the
  Multiphase Interstellar Galactic Halo\altaffilmark{1}}

\author{J. Christopher Howk\altaffilmark{2,3}, 
  Kenneth R. Sembach\altaffilmark{4},
  \& Blair D. Savage\altaffilmark{5}}

\altaffiltext{1}{Based on observations with the NASA/ESA Hubble Space
  Telescope obtained at the Space Telescope Science Institute, which
  is operated by the Association of Universities for Research in
  Astronomy, Incorporated, under NASA contract NAS5-26555.}

\altaffiltext{2}{Center for Astrophysics and Space Sciences,
  University of California, San Diego, C-0424, La Jolla, CA, 92093}

\altaffiltext{3}{Current address: Department of Physics, University of Notre Dame, Notre Dame, IN 46556; jhowk@nd.edu}

\altaffiltext{4}{Space Telescope Science Institute, 
  Baltimore, MD, 21218; sembach@stsci.edu}

\altaffiltext{5}{Astronomy Department, University of Wisconsin-Madison,
  Madison, WI, 53711; savage@astro.wisc.edu}


\begin{abstract}
  
  We describe a new method for accurately determining total gas-phase
  abundances for the Galactic halo interstellar medium with minimal
  ionization uncertainties.  For sight lines toward globular clusters
  containing both ultraviolet-bright stars and radio pulsars, it is
  possible to measure column densities of \HI\ and several ionization
  states of selected metals using ultraviolet absorption line
  measurements and of \HII\ using radio dispersion measurements.  By
  measuring the ionized hydrogen column, we minimize ionization
  uncertainties that plague abundance measurements of Galactic halo
  gas.  We apply this method for the first time to the sight line
  toward the globular cluster Messier 3 [$(l,b)=(42\fdg2, +78\fdg7);
    \ d=10.2 \ {\rm kpc}, z=10.0 \ {\rm kpc}$] using {\em Far
    Ultraviolet Spectroscopic Explorer} and {\em Hubble Space
    Telescope} ultraviolet spectroscopy of the post-asymptotic giant
  branch star von Zeipel 1128 and radio observations by Ransom et al.
  of recently-discovered millisecond pulsars.  The fraction of
  hydrogen associated with ionized gas along this sight line is
  $(45\pm5)\%$, with the warm ($T\sim10^4$ K) and hot ($T\ga10^5$ K)
  ionized phases present in roughly a 5:1 ratio.  This is the highest
  measured fraction of ionized hydrogen along a high-latitude pulsar
  sight line.  We derive total gas-phase abundances $\log N({\rm
    S})/N({\rm H}) = -4.87\pm0.03$ and $\log N({\rm Fe})/N({\rm H}) =
  -5.27\pm0.05$.  Our derived sulfur abundance is in excellent
  agreement with recent solar system determinations of Asplund,
  Grevesse, \& Sauval.  However, it is $-0.14$ dex below the solar
  system abundance typically adopted in studies of the interstellar
  medium.  The iron abundance is $\sim-0.7$ dex below the solar system
  abundance, consistent with the significant incorporation of iron
  into interstellar grains.  Abundance estimates derived by simply
  comparing \sii\ and \feii\ to \HI\ are +0.17 and +0.11 dex higher,
  respectively, than the abundance estimates derived from our refined
  approach.  Ionization corrections to the gas-phase abundances
  measured in the standard way are, therefore, significant compared
  with the measurement uncertainties along this sight line.  The
  systematic uncertainties associated with the uncertain contribution
  to the electron column density from ionized helium could raise these
  abundances by $\la +0.03$ dex ($+7\%$).  Uncertainties in the amount
  of very hot gas ($T\sim10^6$ K) along the line of sight could also
  affect these determinations.

\end{abstract}



\section{Introduction}
\label{sec:intro}

Studies of gas-phase elemental abundances of the diffuse interstellar
medium (ISM) of the Milky Way and other galaxies can give important
information about dust composition, interstellar mixing, and chemical
evolution.  Most absorption line studies of gas associated with the
warm neutral medium (WNM) of the Milky Way and other galaxies derive
gas-phase abundances by comparing columns of the dominant ionization
stage of a metal species in the WNM\footnote{We refer to the dominant
  ionization states of elements in the warm neutral medium as ``low
  ions.''  The dominant ionization stage of an element is typically
  the first whose ionization potential is higher than that of neutral
  hydrogen.  We refer to the next higher ionization states as
  ``intermediate ions.''}  with the column of neutral hydrogen,
assuming (often implicitly) no ionization corrections are needed.
Sight lines through the multiphase ISM of a galaxy not only probe
neutral material, but often include gas associated with the warm
ionized medium (WIM) and hot ionized medium (HIM).  Ionized gas can
represent up to $\sim50\%$ of the total hydrogen column density for
sight lines through the Galactic halo, with an average value of
$\sim25\%$ (Reynolds 1993).  While the WIM likely contains very little
\HI, the dominant ionization states of many metals are the same for
gas in the WNM and the WIM (e.g., \sii, \feii, \ion{Si}{2} -- see
Sembach et al.  2000 and Haffner, Reynolds, \& Tufte 1999).  Thus,
comparing the column densities of the dominant ionization states of
metals to the column density of neutral hydrogen along a sight line
will give apparent metal abundances that are too high compared with
the true values.  The neglect of ionization corrections may be a
source of non-negligible systematic uncertainties to measurements of
interstellar gas-phase abundances (Sembach et al.  2000; Howk \&
Sembach 1999).

In this paper we present a new approach for studying gas-phase
abundances along extended paths through the multiphase halo of the
Milky Way that circumvents such ionization uncertainties.  Our
strategy is to observe ultraviolet (UV) absorption lines from low and
intermediate ions toward UV-bright stars in globular clusters that
contain pulsars.  This allows us to measure all of the ionization
states of certain metals (e.g., sulfur and iron) present in the WNM
and WIM in these directions.  The total hydrogen column to which these
metals are compared is the sum of the \HI\ and \htwo\ column
densities, derived from UV spectra, and the \HII\ column density,
determined using radio observations of the dispersion measures to the
pulsars (which gives the electron column density).  We present a full
discussion of our methodology in \S \ref{sec:method}.

We use this method to study the gas-phase abundances along the sight
line toward the post-asymptotic giant branch star von Zeipel 1128
(\vz) in the distant globular cluster Messier 3 (M~3).  This sight
line probes gas in the first 10 kpc above the Galactic plane.  We
discuss the UV absorption line and pulsar dispersion measure
observations of this sight line in \S \ref{sec:application}.  The
general properties of this interstellar sight line have previously
been discussed by Howk, Sembach, \& Savage (2003; hereafter \pone).
We also apply our new strategy for determining gas-phase abundances in
this section, showing that the neglect of ionization corrections can
lead to significant errors in the derived metal abundances along this
(and presumably other) high-latitude sight lines.  We end with a
general discussion of the implications of this work in \S
\ref{sec:conclusions}.

\section{Methodology}
\label{sec:method}

In a multiphase medium, the total abundance, $A(X)$, of an element $X$
with respect to hydrogen is
\begin{equation}
\label{eqn:simpleabundance}
        A(X) \equiv \frac{\sum\limits_j N(X^j)}
                { N(\mbox{\HI}) + N(\mbox{\HII}) + 2N({\rm H}_2)},
\end{equation}
where $N(X^j)$ is the column density of the $j$th ionization stage of
$X$.  The numerator represents the total column density of the metal
$X$, while the denominator is $N(\mbox{\HI}) + N(\mbox{\HII}) +
2N({\rm H}_2) \equiv N({\rm H})$, the total hydrogen column density.
It is unusual that all the terms in both the numerator and the
denominator can be measured.  One approach is to identify special
sight lines along which this is possible, such as those toward
globular clusters that contain both UV-bright stars and radio pulsars.

The derivation of metal ion column densities is done using UV
absorption line measurements against background sources (in our case
against UV-bright stars) using the {\em Far Ultraviolet Spectroscopic
  Explorer} (\fuse) and/or the UV spectrographs on-board the {\em
  Hubble Space Telescope} (\hst).  A review of UV absorption line
abundance techniques and measurements is given by Savage \& Sembach
(1996).  The transitions accessible to these instruments (with
wavelength ranges of $912 \la \lambda \la 1187$ \AA\ for \fuse\ and
$1150 \la \lambda \la 3100$ \AA\ for \hst) limit the species for which
we can reasonably approximate the total summation in the numerator of
equation (\ref{eqn:simpleabundance}).  Sulfur and iron are the only
elements for which measurements are typically feasible for all of the
ionization states expected to contribute significantly in the WNM and
WIM (see Table 1 of Howk \& Savage 1999 for a summary of WIM tracers).
Measurements of gas associated with the HIM can also be made using
absorption line spectroscopy of the \ovi\ 1031.926 and 1037.617 \AA\ 
doublet.  However, it is not possible to measure the ions of S and Fe
that dominate in the HIM.  For example, the dominant ionization states
of sulfur for $\log T\ga5.5$ are \ion{S}{7} and higher (Sutherland \&
Dopita 1993), which have no measurable UV transitions.  We will make a
correction (described below) for these still-unseen ions when deriving
$A(X)$.

The neutral and molecular hydrogen column densities used in the
denominator of equation (\ref{eqn:simpleabundance}) are also measured
using UV absorption line techniques and instruments.  \HI\ column
densities along sight lines to UV background sources are usually
measured using the strong damping wings of the hydrogen \lyalpha\ 
transition at 1215.670 \AA\ (e.g., Jenkins 1971; Bohlin, Savage, \&
Drake 1978; Diplas \& Savage 1994).  Measurements of \htwo\ are made
using absorption lines of the Lyman and Werner electronic transitions
in the \fuse\ bandpass (e.g., Savage et al.  1977; Shull et al. 2000).

We derive the \HII\ column density from the radio dispersion measure,
$DM = \Nelectron \equiv \int n_e dl$, observed toward pulsars in the
target globular clusters.  We make two corrections to the observed
\Nelectron\ in order to derive $N(\mbox{\HII})$ from \Nelectron: (1) a
correction for the contribution to the electrons from ionized helium
(the contribution from heavier elements is negligible); and (2) the
removal of the contribution of the HIM, for which metal ions are
typically not measurable.

The electron column in H+He gas is $\Nelectron = N(\mbox{\HII}) +
N(\mbox{\ion{He}{2}})+ 2N(\mbox{\ion{He}{3}})$.  We define the helium
correction factor $\eta$ such that $N(\mbox{\HII}) = \eta
N(\mbox{\electron})$.  Therefore,
\begin{equation}
\label{eqn:eta}
        \eta \equiv \frac{1}{1+\abun{He}x({\rm He}^+)/x({\rm H}^+)+ 
                2\abun{He}x({\rm He}^{+2})/x({\rm H}^+)},
\end{equation} 
where the ionization fraction of an ion $X^j$ in the ionized gas is
$x(X^j)\equiv N(X^j)/N(X)$.  We assume $\abun{He}=0.1$ throughout.
Values for $\eta$ will typically be in the range 0.8 to 1.0.

Emission line studies of the WIM indicate $0.67 \la x({\rm
  H^+})\subwim \la 1.0$ (Reynolds 1989; Reynolds et al.  1998) and
$x({\rm He^+})\subwim \la 0.27 \, x({\rm H^+})\subwim$ (Reynolds \&
Tufte 1995).  Most photoionization models of the WIM predict $x({\rm
  He^{+2}})\subwim$ to be very small (Sembach et al.  2000; Mathis
2000), giving values of $\eta\subwim$ near 1.0.  However, Arabadjis \&
Bregman (1999) suggest $x({\rm He^{+2}})\subwim$ is non-negligible on
the basis of X-ray absorption estimates (although see Slavin, McKee,
\& Hollenbach 2000 for alternative interpretations of these data).  In
what follows, we consider two cases: $x({\rm He}^{+2})\subwim \approx
0$ and $x({\rm He}^{+2})\subwim = 1-x({\rm He^+})\subwim$; we refer to
these as the minimum and maximum helium ionization cases,
respectively, and tabulate characteristic values of $\eta$ in Table
\ref{tab:eta}.  We favor the minimum helium ionization model given
that much of the sulfur in the WIM is in the form of \sii\ (Haffner et
al.  1999), implying the ionization state of metals in the WIM is
relatively low.

It is extremely difficult to measure the dominant metal ionization
states in the HIM.  We therefore subtract an estimated HIM
contribution, \Nelectron\subhim, from \Nelectron\ to derive a value of
$N(\mbox{\HII})$ appropriate for the WIM alone.  We use measurements
of \ovi\ to estimate \Nelectron\subhim, adopting the following
two-component prescription:
\begin{equation}
%
%
\Nelectron\subhim  = 
    \frac{1}{\eta\subhim} 
    \left[ \frac{N(\mbox{\ovi})}{A({\rm O}) x({\rm O}^{+5})}        
   + \langle n_{\rm H} \rangle_c \, d \right],
\label{eqn:himelectrons1}
\end{equation}
where $N(\mbox{\ovi})$ is the measured \ovi\ column density, and $d$
is the distance of the globular cluster from the sun.  The first term
in brackets is the column density of $T\sim3\times10^5$ K gas
associated with the thick disk distribution of the Milky Way, which is
best probed by \ovi\ (Wakker et al.  2003; Savage et al.  2003).  The
second term accounts for an extended hot corona assumed to be
uniformly distributed and characterized by $T\ga10^6$ K and a total
hydrogen density $\langle n_{\rm H} \rangle_{c} \sim 2\times\e{-4}$
\percc.  The existence of this component is derived from observations
of \ovi\ associated with high-velocity clouds, which suggest these
clouds are interacting with a hot, low-density plasma extending to
large distances ($>70$ kpc) from the Sun (Sembach et al.  2003).
Supporting evidence for a hot corona is provided by very strong zero
redshift X-ray absorption lines of \ion{Ne}{9}, \ion{O}{7} and
\ion{O}{8} in the spectra of X-ray bright active galaxies (Nicastro et
al. 2002; Fang et al. 2003; Rasmussen, Kahn, \& Paerels 2003) and
Galactic X-ray sources (Futamoto et al. 2004; Yao \& Wang 2005).


Defining the WIM electron column as $\Nelectron \subwim = \Nelectron -
\Nelectron \subhim$, i.e., correcting for the contribution from the
HIM, we rewrite Eqn. (\ref{eqn:simpleabundance}) as
\begin{equation}
\label{eqn:complexabundance}
        A(X) = \frac{\sum\limits_{j(WNM,WIM)} N(X^j)}
                {[N(\mbox{\HI})+
                  \eta\subwim N(\mbox{\electron})\subwim+
                  2N({\rm H}_2)]},
\end{equation}
noting the summation is over metal ionization states found in the WNM
and WIM (i.e., excluding the HIM).  We suggest using numerical values
for the minimum and maximum helium ionization cases of $\eta\subwim =
0.98\pm0.01$ and $0.81\pm0.04$, where the uncertainties correspond to
the dispersion of values found in Table \ref{tab:eta}.

The differences in the helium ionization assumptions represent the
dominant source of systematic uncertainty in our method.  The
magnitude of the uncertainty is roughly $0.2 \times
N(\mbox{\HII})\subwim/N({\rm H})$, or $\la10\%$ for high-latitude
sight lines.  While spatial variations of $N(\mbox{\electron})$
between the sight lines to the M~3 pulsars and the UV bright stars
could be another potential source of systematic uncertainty, it seems
in most cases this effect is negligible.  For example, Freire et al.
(2001) find the dispersion in the $DM$ values toward 15 pulsars in 47
Tuc is only $0.5\%$ of the average with a full range of $\approx2\%$
of the average (for pulsars with a maximum angular separation of
$\approx2\arcmin$).

\section{Application to the Sight Line Toward von Zeipel 1128 in Messier 3}
\label{sec:application}

We now apply the approach outlined above to determine average
gas-phase abundances of sulfur and iron in the WNM and WIM toward the
high-latitude globular cluster M~3 [$(l,b)=(42\fdg2, +78\fdg7); \ 
d=10.2 \ {\rm kpc}, z=10.0 \ {\rm kpc}$].  We describe in the
following subsections the UV data and analysis that are used to derive
the metal and neutral hydrogen column densities, the radio
observations used to derive the electron column density toward M~3,
and the resulting analysis of the total abundances of S and Fe in this
direction.

\subsection{UV Absorption Line Measurements of Metals and Neutral Hydrogen}

In \pone\ we presented UV absorption line measurements of the ISM in
the direction of \vz.  That work was based upon \fuse\ observations
covering the spectral range 905--1187 \AA\ with a resolution of
$\sim20$ \kms.  In this work we present new observations of \vz\ 
obtained with the Space Telescope Imaging Spectrograph (STIS) on board
\hst\ under guest observer proposal GO 9150.  Proffitt et al.  (2001)
discuss the detailed characteristics of this instrument.

The properties of our STIS observations of \vz\ are summarized in
Table \ref{tab:observations}.  A total of 12.7 ksec of exposure time
was collected with each of the E140M and E230M intermediate resolution
echelle gratings.  These data have signal-to-noise ratios varying from
$\sim10$ to 20 per resolution element. The shorter wavelength E140M
grating observations have a resolution of $\sim6.5$ \kms\ (FWHM),
while the longer wavelength E230M observations have a resolution
$\sim10$ \kms\ (FWHM).  All observations were made through the
$0\farcs2\times 0\farcs06$ apertures.  The STIS data were calibrated
using v2.17 of the CALSTIS pipeline.

We use the STIS data to derive column densities and limits for the
species \HI, \ion{S}{1}, \sii, and \ion{Fe}{1}.  The details of our
measurements of metal line and \HI\ column densities are presented
below.  We do not make use of the longest-wavelength E230M exposure
(archive ID O6F501030).  An analysis of the full STIS spectrum will be
presented in a future work.

\subsubsection{\HI\ Column Density from \lyalpha}

We derive the interstellar \HI\ column density of the \vz\ sight line
by fitting the damping wings of the \lyalpha\ profile.  We generally
follow the procedures described by Sonneborn et al.  (2002) for
determining the \HI\ column density and uncertainty.

\begin{figure}
\epsscale{0.9}
\plotone{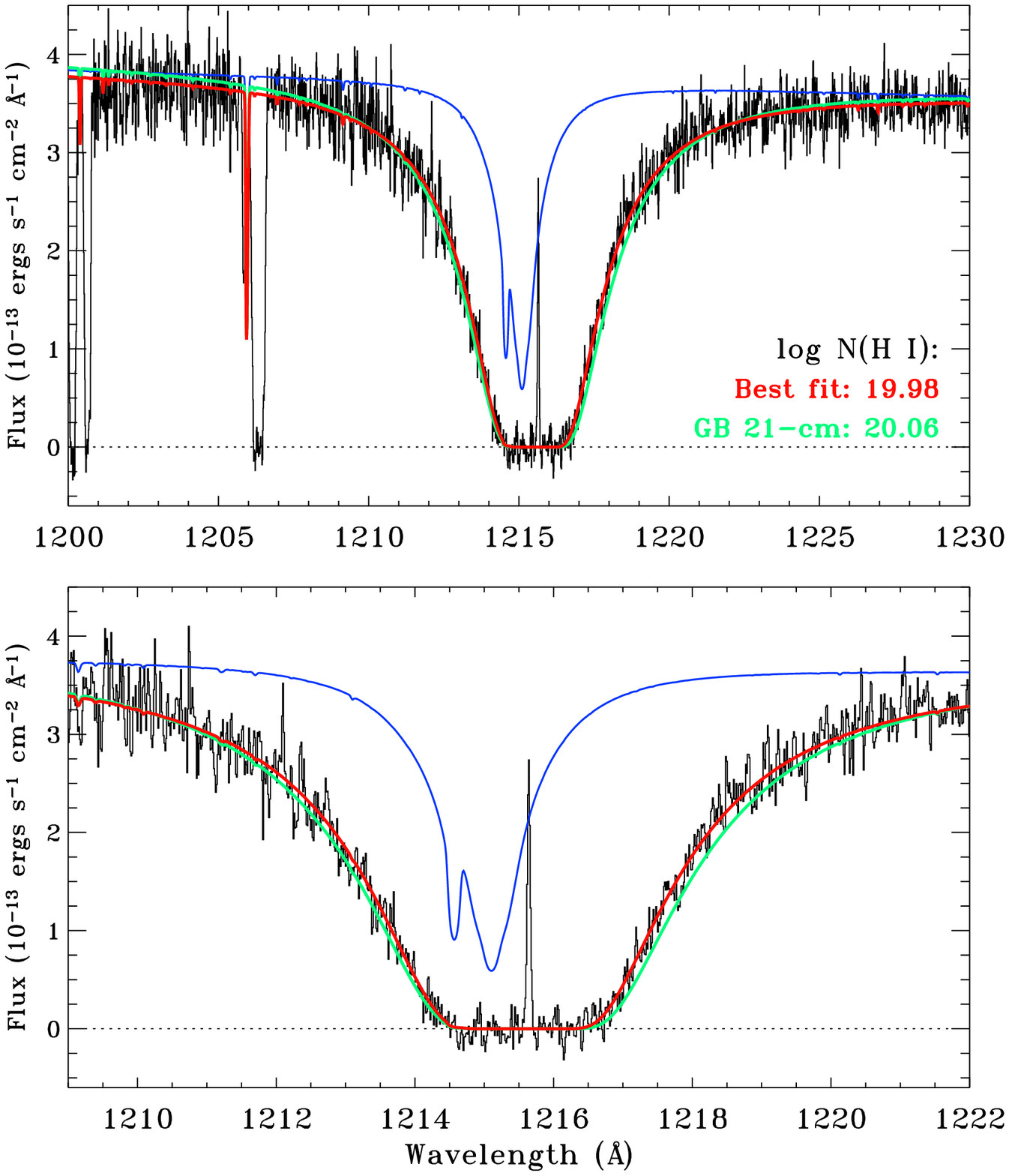}
\caption{Two views of the STIS E140M spectrum of the \lyalpha\ 
  absorption profile toward \vz.  Both stellar and interstellar
  absorption contribute to this profile.  The estimated stellar
  profile is shown as the thin blue line and has been shifted to
  $v_{LSR}=-145$ \kms\ to match the observed positions of stellar
  absorption lines in the STIS spectrum.  The best fit to the
  interstellar and stellar absorption profile is shown in red,
  corresponding to an interstellar column density $\log N(\mbox{\HI})
  = 19.98\pm0.03$.  The green line shows the best fit profile
  adopting the \HI\ column density derived from Green Bank 140-ft
  telescope observations of 21-cm emission in this direction, $\log
  N(\mbox{\HI}) = 20.04\pm0.01$ (see \pone).  The sharp emission line
  in the center of the \lyalpha\ absorption trough is geocoronal
  emission.  The strong stellar line near 1206 \AA\ is stellar
  \ion{Si}{3}.
  \label{fig:lyalpha}}
  

%
\end{figure}

Figure \ref{fig:lyalpha} shows the STIS spectrum of the \lyalpha\ 
absorption line toward \vz.  This absorption line contains
contributions from both the stellar atmosphere of \vz\ and the ISM in
this direction.  We use a model stellar atmosphere to normalize the
data during the fitting process.  Our adopted atmosphere, kindly
provided by P. Chayer (2002, private communication), was calculated
using the SYNSPEC package of I.  Hubeny (2000, private communication)
and assumed $T_{eff} = 35,000$ K, $\log g = 4.0$, and He/H$\,=0.1$
(Dixon, Davidsen, \& Ferguson 1994).  The adopted atmosphere is shown
as the blue line in Figure \ref{fig:lyalpha}.  We adopt a second-order
Legendre polynomial correction to the model atmosphere to match it to
the STIS spectrum, following Sonneborn et al. (2002).  This is done to
account for any calibration and order combination uncertainties in the
data or any uncertainties in the large-scale flux distribution of the
models.  The parameters of the polynomial are allowed to vary during
the fitting process.

The interstellar column density that best fits the STIS observations
of \lyalpha\ is $\log N(\mbox{\HI}) = 19.98\pm0.03$.  This is slightly
different (0.01 dex higher) than the value quoted in \pone\ and is due
to a better calibration and combination of the data.  The best-fit
profile is shown as the red line in Figure \ref{fig:lyalpha}.  Our
error estimate includes contributions from the uncertainties in the
adopted properties of the stellar model (c.f., Sonneborn et al.
2002), which dominate the total error budget.

Two observations of the \HI\ 21 cm emission line in this direction
give slightly higher column density estimates (as discussed in \pone).
Danly et al. (1992) collected \HI\ 21 cm emission line observations
toward \vz\ using the NRAO Green Bank 140-ft. telescope, which has a
beam of $21\arcmin$.  The Leiden-Dwingeloo Survey (LDS; Hartmann \&
Burton 1997), which has a beam size of $35\arcmin$, includes a
pointing centered $\approx12\arcmin$ from \vz.  Table \ref{tab:hi}
gives the \HI\ column densities in the direction of M 3/\vz\ derived
from the two 21-cm observations and the \lyalpha\ profile.  The NRAO
and first LDS value were derived here through a direct integration of
the brightness temperature distribution over the range $-75 \le \mvlsr
\le +75$ \kms\ in the optically-thin limit (optical depth effects are
$<1\%$ for this high-latitude sight line).  The second LDS \HI\ value
is from Wakker et al.  (2003); it is the sum of three Gaussians fit to
a weighted average of four LDS pointings in this direction.  The
difference in column densities between these two treatments of the LDS
data is likely due to the slightly different spectra adopted (a single
pointing versus a weighted average of four pointings) and the
different techniques used to derive column densities (a direct
integration versus Gaussian fits).  Both the Danly et al. (1992) and
LDS observations have been corrected for stray radiation. The
uncertainties given in Table \ref{tab:hi} do not include contributions
from errors in this correction.  Lockman \& Savage (1995) estimate the
stray radiation correction uncertainty of observations similar to
those of Danly et al. (1992) to be of order $\sigma[ N(\mbox{\HI})]
\sim10^{19}$ \column, or $\sim0.04$ dex for the column densities in
Table \ref{tab:hi}.

All of the \HI\ columns derived using 21-cm observations are higher
than that derived above via a direct fit to \lyalpha\ absorption
toward \vz.  The largest difference is 0.08 dex, or $20\%$, between
the NRAO spectrum and the fit to \lyalpha.  This difference is likely
due to structure in the \HI\ column on scales smaller than the 21 cm
beam (see Savage et al. 2000), although it could also be caused by
systematic effects (e.g., unaccounted for uncertainties in the adopted
stellar \lyalpha\ profile or uncertainties in the 21 cm stray
radiation correction).  In Figure \ref{fig:lyalpha} we show the
best-fit \lyalpha\ profile if we assume $\log N(\mbox{\HI})=20.06$ as
the green line.  The polynomial continuum correction is allowed to
vary when fitting this profile to the data.  This column over-predicts
the optical depth in the wings of \lyalpha.  In what follows, we
assume the \lyalpha\ profile provides the best estimate of
$N(\mbox{\HI})$ in this direction.

\subsubsection{Metal Line Column Densities}

We have used the new STIS data to derive column densities and limits
for the metal species \ion{S}{1}, \sii, and \ion{Fe}{1}.  Our
measurements of the metal line properties toward \vz\ follow \pone. We
have fit the stellar continuum in the regions surrounding metal
transitions using low order Legendre polynomials.  Following Sembach
\& Savage (1992), we have directly integrated the observed line
profiles and apparent optical depths to arrive at the equivalent
widths and apparent column densities, $N_a$ (Savage \& Sembach 1992).
These measurements are presented in Table \ref{tab:measurements}.  We
also list our adopted central wavelengths and oscillator strengths
(all from Morton 2003) for the transitions, the velocity range over
which the interstellar absorption lines were integrated, and an
empirical estimate of the signal to noise ratio per resolution element
in the region about each line.  The quoted measurement errors include
contributions from continuum placement uncertainties and the effects
of a 2\% error in the flux zero level (c.f., Sembach \& Savage 1992).

\begin{figure}
  \epsscale{0.6} \plotone{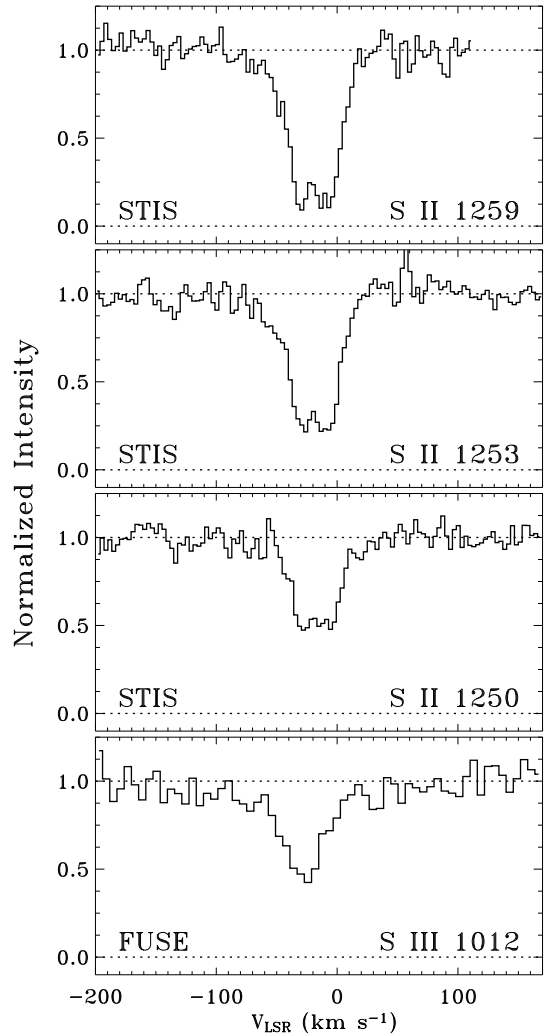}
\caption{Absorption line profiles of \ion{S}{2} $\lambda 1250.584, 
  1253.811$, and 1259.519 from STIS E140M observations and \ion{S}{3}
  $\lambda1012.495$ from {\em FUSE} (see Paper~I).  We suggested in
  \pone\ that the material at $\mvlsr \approx -30$ km s$^{-1}$ was
  likely associated with the Galactic thick disk given the higher
  degree of ionization compared with the material at $\mvlsr \approx
  -4$ km s$^{-1}$.  The STIS data have a resolution of $\sim6.5$ km
  s$^{-1}$, while the {\em FUSE} data have a resolution $\sim20$ km
  s$^{-1}$.
\label{fig:profiles}}
\end{figure}

Of the metal species considered in this work, only \sii\ is detected
in the STIS data.  The absorption line profiles of the \sii\ triplet
at 1250.584, 1253.811, and 1259.519 \AA\ from STIS E140M observations
of \vz\ are shown in Figure \ref{fig:profiles}; also shown is the
profile of the \sthree\ 1012.495 \AA\ line from the \fuse\ 
observations of this star (\pone).  To calculate the limiting
equivalent widths and apparent column densities of \ion{S}{1} and
\ion{Fe}{1}, we assume the gas traced by these species has an
intrinsic width of $\sim40$ \kms\ (FWHM), which is the value derived
from a single Gaussian profile fit to the \sii\ 1250.584 \AA\ 
transition.  All limits given in this work are $3\sigma$.

The integrated apparent column densities of the weaker two transitions
at 1250.584 and 1253.811 \AA, which have $f$-values that are different
by a factor of two, differ by $\sim0.01$ dex, with the weaker
transition giving a higher $N_a$.  The strongest line, with an
$f$-value three times that of the weakest, gives a value of $N_a$ that
is -0.05 dex lower than the weakest line.  The progression of
decreasing $N_a$ with increasing $f$ suggests these lines may contain
unresolved saturated structure (Savage \& Sembach 1991).

\begin{figure}
\epsscale{0.6}
\plotone{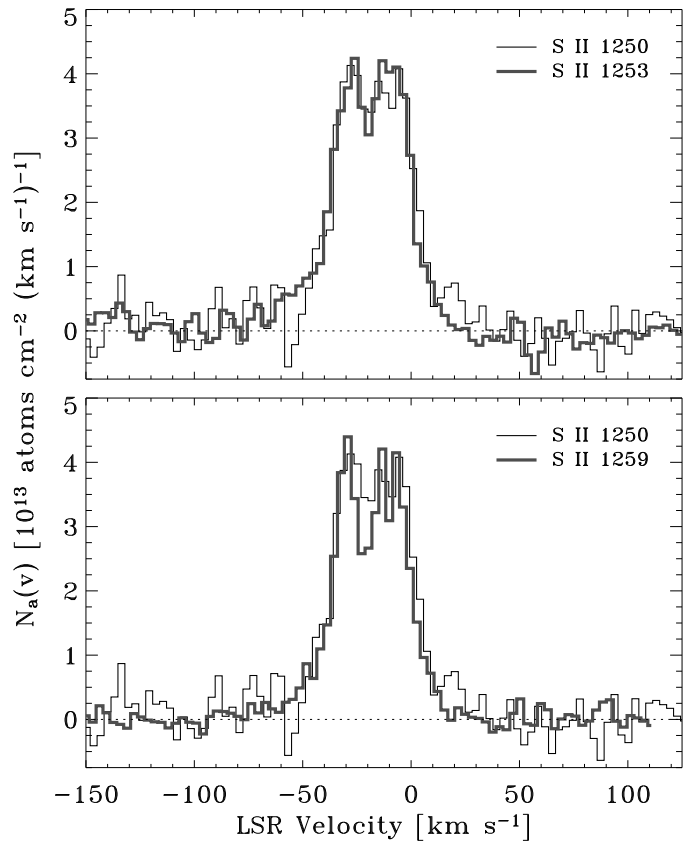}
\caption{Comparison of the apparent column density profiles for \ion{S}{2} 
  $\lambda 1250.584, 1253.811$, and 1259.519 from STIS E140M
  observations.  The weaker two lines (at 1250.584 and 1253.811 \AA)
  are compared in the top panel; the strongest line (1259.519 \AA) is
  compared with the weakest in the lower panel.  The lack of agreement
  between these profiles indicates the stronger line contains a small
  amount of unresolved saturated structure.
\label{fig:nav}}
\end{figure}

Figure \ref{fig:nav} shows a comparison of the apparent column density
profiles as a function of velocity, $N_a(v)$, for the three
transitions.  In each of the panels, the weakest \sii\ transition at
1250.584 \AA\ is shown as the thin black histogram, while the two
stronger transitions are plotted as thick gray histograms.  The
$N_a(v)$ profiles in Figure \ref{fig:nav} do not show the classical
symptoms of unresolved saturated structure, with the highest peaks of
the strong line profiles suppressed compared with the weak line
profiles.  The central dip near $\mvlsr\approx-20$ \kms\ in the \sii\ 
1259 \AA\ profile may be indicating the presence of a blended, but
saturated, component at these velocities.

Because of the possible presence of saturated structure in these
profiles, we apply a correction to the data following the methodology
outlined in Savage \& Sembach (1991).  We derive this correction
separately for the line pairs $\lambda 1250$+$\lambda 1253$ and
$\lambda 1253$+$\lambda 1259$ and average the results.  Our adopted
total column density is $\log N(\mbox{\sii}) = 15.28\pm0.02$.
Independent support for this column density comes from a curve of
growth analysis.  A single component curve of growth fit to the
equivalent widths of these three transitions gives $\log
N(\mbox{\sii}) = 15.28^{+0.03}_{-0.02}$ for a Doppler parameter of
$b=19.5\pm1.2$ \kms.

Table \ref{tab:columns} lists the final column densities adopted in
this work, including those of \sthree, \feii, \fethree, and \ovi\ and
of the limits to \htwo, \ion{S}{4}, and \ion{S}{6} from our analysis
of the \fuse\ data in \pone.

\subsection{Pulsar Dispersion Measurements and the Electron Column
  Density}

\subsubsection{Radio Pulsar Dispersion Measurements}

The electron column density toward M~3 is derived from the dispersion
measures determined by Ransom et al.  (2004) for three binary
millisecond pulsars in M~3.  These pulsars were discovered with the
Arecibo 305-m radio telescope at 20 cm using the Wideband Arecibo
Pulsar Processors and the search algorithms described by Ransom,
Cordes, \& Eikenberry (2003) and Ransom, Eikenberry, \& Middleditch
(2002).

The average dispersion measure toward the three pulsars M~3A, M~3B,
and M~3D is $DM = 26.33\pm 0.15$ pc \percc\ (standard deviation).  We
do not include the unconfirmed pulsar M~3C in this average, although
it gives a consistent $DM$ (Ransom et al.  2004).  Typical
uncertainties in the individual measurements are $\sim0.1$ pc \percc.
The average dispersion measure corresponds to an electron column
density of $\log \Nelectron = 19.912\pm0.002$.  This is the total
electron column, including contributions from the WIM and the HIM
along this sight line.

\subsubsection{Correcting for the Hot ISM}

Applying equation (\ref{eqn:himelectrons1}) to the 10 kpc sight line
toward \vz\ (\pone), which is very close to the north Galactic pole,
yields $\Nelectron\subhim\sim1.2\times10^{19}$ \column.  This estimate
assumes an interstellar gas-phase abundance $A({\rm
  O})=4.08\times\e{-4}$ (Andr\'{e} et al.  2003), an ionization
fraction $x({\rm O}^{+5})=0.2$ (Sutherland \& Dopita 1993; see Savage
et al. 2003), and an \ovi\ column $N(\mbox{\ovi}) = 3.1\times\e{14}$
\column\ (\pone).  We adopt $\langle n_{\rm H} \rangle_{c} \sim
2\times\e{-4}$ \percc\ as the coronal density for the second term in
brackets in equation (\ref{eqn:himelectrons1}).  We have assumed a
medium with fully-ionized hydrogen and helium, $\eta\subhim=0.83$.

We will adopt a 50\% uncertainty in calculations of the HIM column
using the parameters for equation (\ref{eqn:himelectrons1}) adopted
above.  We have assumed $x({\rm O}^{+5})$ equal to its maximum in
collisional ionization equilibrium models (Sutherland \& Dopita 1993).
While much of the \ovi\ may in fact reside in regions where $x({\rm
  O}^{+5})\approx0.2$, the \ovi\ ionization fraction in the transition
temperature gas along extended sight lines through the thick disk of
the Galaxy is uncertain.  We've assumed a single, constant density
component for the $10^6$ K material in an effort to approximate a
Galactic corona.  Yao \& Wang (2005) have estimated the density
distribution of this hot component of the ISM using X-ray absorption
measurements of \ion{Ne}{9}, \ion{O}{7} and \ion{O}{8} toward several
Galactic targets, mostly in the Galactic Center region.  They present
two model density distributions based on their data: a plane-parallel,
exponential disk distribution and a spherical distribution about the
Galactic Center.  These models give estimates of $\Nelectron\subhim
\sim2.9\times\e{19}$ \column\ and $\sim1.2\times\e{19}$ \column,
respectively, for the sight line to \vz\ (with uncertainties $>30$ to
50\%).  However, these results are likely to be strongly influenced by
hot gas in the inner Galaxy.  It is not clear that the Yao \& Wang
density distributions are appropriate for the hot ISM at the solar
circle.  We will adopt the predictions of equation
(\ref{eqn:himelectrons1}) as calculated in the preceding paragraph for
the HIM column and assume a 50\% uncertainty.

\subsection{Total Abundances}

Table \ref{tab:results} summarizes the results of our analysis of the
M~3 sight line, following the method described in \S \ref{sec:method}.
The total column density of sulfur toward \vz\ is $\log N({\rm S}) =
15.34\pm0.02$; the contributions from \ion{S}{1}, \ion{S}{4} and
\ion{S}{6} are negligible [with combined $3\sigma$ limits less than a
few percent of $N({\rm S})$].  The total column of iron is $\log
N({\rm Fe}) = 14.95\pm0.04$, where we have assumed the contributions
from \ion{Fe}{4} and higher ionization states in WNM and WIM gas are
negligible (as they are for ions of sulfur).  

In Table \ref{tab:results} we include the values of \abun{S} and
\abun{Fe} for both the minimum and maximum helium ionization
assumptions.  We give the values adopted for the helium correction,
$\eta\subwim$ and the values of \Nelectron\ associated with the WIM
and the HIM.  Also included in Table \ref{tab:results} are the column
of hydrogen (neutral+ionized) and the fraction of hydrogen found in
the ionized phase.  Those values with a ``Total'' subscript refer to
measurements that include the contribution from the HIM.  Thus, the
hydrogen column $N({\rm H})_{Total}$ includes contributions from the
WNM, WIM, and HIM, whereas $N({\rm H})$ includes contributions from
the WNM and WIM only.  As stated above, the reason for this
distinction is that we have not measured metals associated with the
hot phase with the exception of \ovi.

\section{Discussion and Conclusions}
\label{sec:conclusions}

We have derived the average gas-phase interstellar abundances in a way
that minimizes ionization uncertainties along the sight line to the
halo globular cluster M~3.  The abundance of sulfur toward M~3 is
$\log A(S)=-4.87\pm0.03$.  Because sulfur is not believed to be
significantly incorporated into dust grains, \abun{S} represents the
total average metal abundance of the neutral and ionized ISM along
this sight line.  The sight line to M~3 probes warm material
associated with the thin interstellar disk and the extended (thick
disk) distribution of neutral and ionized gas (\pone).

Typically, the solar system abundance for sulfur adopted by ISM
studies is $\log \abun{S}_\odot=-4.73$ (Savage \& Sembach 1996), an
average of the discrepant photospheric, $\log \abun{S}_\odot =
-4.67\pm 0.11$, and meteoritic, $\log \abun{S}_\odot=-4.80\pm 0.06$,
values from Grevesse \& Sauval (1998).  Thus, our total sulfur
abundance along the sight line to M~3 is $-0.14\pm0.03$ dex below, or
$\sim3/4$ of, the commonly-adopted solar value.\footnote{In the
  remaining discussion we adopt the results of the minimum helium
  ionization case, our preferred model.  For simplicity we do not
  explicitly refer to the $\sim 0.03$ dex systematic uncertainty
  associated with this choice of model.}  Recent revisions to the
solar composition, however, are in very good agreement with our new
measurement.  Asplund, Grevesse, \& Sauval (2005) derive $\log
\abun{S}_\odot = -4.86\pm 0.05$ for the solar photosphere, while
incorporating the meteoritic data from Lodders (2003) with their
results gives a meteoritic abundance of $\log \abun{S}_\odot =
-4.84\pm 0.04$.\footnote{Because H is deficient in meteorites, the
  metal abundances in meteorites are derived relative to Si.  The
  absolute meteoritic abundances are then calculated from an adopted
  solar photospheric Si/H ratio.  In this case, Asplund et al. (2005)
  use $\log \abun{Si}_\odot = 7.51\pm0.04$.} It is somewhat
disconcerting for ISM studies that the recommended solar abundances
are in such a state of flux.  However, the excellent agreement between
our determination of $A(S)$ toward \vz\ and the most recent solar
system estimates is encouraging.

In contrast to sulfur, iron is heavily incorporated into grains.  Our
iron abundance measurement is $\log \abun{Fe} = -5.26\pm0.04$.  Most
of the difference between the gas-phase interstellar abundance and the
solar system abundance, $\log \abun{Fe}_\odot = -4.55\pm0.03$ (Asplund
et al. 2005), is due to the incorporation of Fe into dust in both the
WNM and WIM.


The ratio of sulfur to hydrogen derived in the usual way toward
\vz\ is $\log N(\mbox{\sii})/N(\mbox{\HI}) = -4.70\pm0.04$, consistent
with the Grevesse \& Sauval (1998) solar system value and $+0.17$ dex
higher than the total abundance derived above using our method of
correcting for ionized gas contributions.  The neglect of ionization
corrections causes a systematic error that is significantly higher
than the estimated statistical uncertainties.  Similarly, we derive
$\log N(\mbox{\feii})/N(\mbox{\HI}) = -5.18\pm0.06$, implying a
systematic error of $+0.09$ dex.  The abundances derived by comparing
singly-ionized sulfur and iron with \HI\ both overestimate the true
gas-phase abundances by significant amounts.  We note that although
the direct comparison of two singly-ionized species reduces the impact
of ionization uncertainties, they are still not negligible compared
with typical statistical uncertainties.  For the sight line toward M~3
we find $\log N({\rm Fe})/N({\rm S}) = -0.40\pm0.05$, while simply
comparing the column densities of singly-ionized atoms gives $\log
N(\mbox{\feii}) / N(\mbox{\sii}) = -0.48\pm0.05$.  For comparison, the
solar system ratio from Asplund et al. (2005) is $\log N({\rm
  Fe})/N({\rm S}) = +0.31$, while the standardly-adopted value in ISM
studies has been +0.23 (see Savage \& Sembach 1996).

The overestimate of gas-phase abundances derived by comparing
singly-ionized species with neutral hydrogen is not unique to this
sight line.  While the fraction $N(\mbox{\HII})_{Total}/N({\rm
  H})_{Total} \approx 0.45\pm0.04$ for the \vz\ sight line is the
highest yet reported (Reynolds 1993), the WIM will contribute to the
singly-ionized metal column densities along all high-latitude sight
lines.  Several groups have derived near-solar abundances of sulfur
comparing \sii\ with \HI\ along high-latitude sight lines with \HI\ 
column densities similar to that toward \vz\ in M~3 (e.g., Howk et al.
1999; Spitzer \& Fitzpatrick 1993).  It is likely that the abundances
derived in these studies suffer from the systematic effects of ionized
gas as well.

It may be possible to model the contribution from the ionized gas
along a sight line and correct for its systematic effects on abundance
studies.  Using the WIM models constructed by Domg\"{o}rgen \& Mathis
(1994), Sembach \& Savage (1996) estimated that the magnitude of
ionization uncertainties for derived gas-phase elemental abundances
should be less than $\sim0.1$ to 0.2 dex along most sight lines
through the Galactic halo, consistent with the values derived in this
work.

Unless $\abun{S}$ varies significantly within the Galaxy, however,
correcting for ionization effects may not be so straightforward.  Howk
et al. (1999), in their detailed study of the sight line to $\mu$
Columbae, derived $\log N(\mbox{\sii})/N(\mbox{\HI}) = -4.65\pm0.02$;
they used photoionization models to argue that the sulfur abundance
along this sight line was near their adopted solar system value of
$\log \abun{S}_\odot = -4.73$ (as recommended by Savage \& Sembach
1996).  In this case, the star is much closer to the Sun than \vz, and
Howk et al. assumed an \HII\ region about the star dominated the
ionized gas content of the sight line.  Even adopting the maximum
ionization correction derived by Howk et al. (1999), however, gives a
sulfur abundance $\approx 0.2$ dex higher than that derived toward
\vz\ in this work.

Sembach et al. (2000) produced detailed models for the ionization of
the WIM by OB stars and used the models to estimate the ionization
corrections required for the sight line to the distant halo star
HD~93521 (Spitzer \& Fitzpatrick 1993).  Their models predict an
ionization correction of $\approx-0.2$ dex to \abun{S} derived by
Spitzer \& Fitzpatrick (1993), giving \abun{S} consistent with that
derived in this work.  However, applying their method to the sight
line toward \vz\ yields an abundance $\log \abun{S} = -5.3$, or
$\approx-0.4$ dex from the value derived in this work.

These examples call into question our ability to effectively model the
effects of WIM gas more generally.  Indeed, models designed to match
emission line diagnostics may not be good descriptors of the
absorption line diagnostics of the WIM, which are more sensitive to
low-density gas.

We can state with confidence that abundance studies of almost all
high-latitude, low hydrogen column density [$\log N(\mbox{\HI}) \la
20$] sight lines are affected by the presence of ionized gas when
comparing singly-ionized metal species with \HI.  The degree to which
an individual sight line is contaminated likely depends on the
fraction of hydrogen along the sight line that is ionized.  The work
of Reynolds (1993) suggests that this fraction can range from
$\sim20\%$ to $\sim50\%$.  We note that some species can be reliably
compared if they have similar ionization properties; for example, the
ratio \ion{O}{1}/\HI\ should not be affected by ionization effects
like those discussed here given the strong charge exchange reaction
that locks the ionization state of these species together.
Unfortunately, the far-UV \ion{O}{1} lines toward \vz\ are strongly
saturated while the weak \ion{O}{1} 1355 \AA\ transition is not
detected.

We will apply the technique outlined here to more sight lines and to
an extended array of metal species in the future to study this source
of systematic uncertainties in regions of varying physical conditions
and fractional ionization.  At least three other globular clusters
contain both radio pulsars and stars with sufficient UV flux to allow
measurements like those presented in this work.

\acknowledgements

We thank S. Ransom for discussions regarding the dispersion measures
toward pulsars in M~3.  Support for proposal numbers HST-GO-9150.03-A
and HST-GO-9410.03-A was provided by NASA through grants from the
Space Telescope Science Institute, which is operated by the
Association of Universities for Research in Astronomy, Incorporated,
under NASA contract NAS5-26555.  K.R.S. acknowledges support from NASA
contract NAS5-32985 to the Space Telescope Science Institute.




\begin{deluxetable}{cccc}
\tablenum{1}
\tablecolumns{4}
\tablewidth{0pt}
\tablecaption{Values of Helium Correction Factor $\eta$
\label{tab:eta}}
\tablehead{
\colhead{$x({\rm H}^+)$} & 
\colhead{$x({\rm He}^+)/x({\rm H}^+)$} & 
\colhead{$x({\rm He}^{+2})$} & 
\colhead{$\eta$} 
}
\startdata
\cutinhead{Minimum Helium Ionization\tablenotemark{a}}
0.67 & 0.15 & 0.00 & 0.99 \\
0.80 & 0.15 & 0.00 & 0.99 \\
1.00 & 0.15 & 0.00 & 0.99 \\
0.67 & 0.27 & 0.00 & 0.97 \\
0.80 & 0.27 & 0.00 & 0.97 \\
1.00 & 0.27 & 0.00 & 0.97 \\
\cutinhead{Maximum Helium Ionization\tablenotemark{b}}
0.67 & 0.00 & 1.00 & 0.77 \\
0.80 & 0.00 & 1.00 & 0.80 \\
1.00 & 0.00 & 1.00 & 0.83 \tablenotemark{c}\\
0.67 & 0.15 & 0.78 & 0.80 \\
0.80 & 0.15 & 0.81 & 0.82 \\
1.00 & 0.15 & 0.85 & 0.84 \\
0.67 & 0.27 & 0.60 & 0.83 \\
0.80 & 0.27 & 0.66 & 0.84 \\
1.00 & 0.27 & 0.73 & 0.85 \\
\enddata
\tablenotetext{a}{The minimum helium ionization case assumes $x({\rm
    He}^{+2})=0$.}
\tablenotetext{b}{The maximum helium ionization case assumes $x({\rm
    He}^{+2})=1-x({\rm He}^{+})$.}
\tablenotetext{c}{This case is appropriate for HIM gas.}
\end{deluxetable}


\begin{deluxetable}{lccccc}
\tablenum{2}
\tablecolumns{6}
\tablewidth{0pt}
\tablecaption{STIS Observations of vZ 1128\tablenotemark{a}
\label{tab:observations}}
\tablehead{
\colhead{Archive ID} & 
\colhead{Grating} &
\colhead{$t_exp$ [ksec]} &
\colhead{$R\equiv\lambda/\Delta \lambda$} & 
\colhead{FWHM [km/s]} &
\colhead{$\lambda$ Range [\AA]} 
}
\startdata
O6F502010 & E140M & 2.0  & 45000 & 6.5 & 1140--1735 \\
O6F502020 & E140M & 10.7 & 45000 & 6.5 & 1140--1735 \\
O6F501010 & E230M & 2.0  & 30000 & 10 & 1575--2380 \\
O6F501020 & E230M & 8.0  & 30000 & 10 & 1575--2380 \\
O6F501030 & E230M & 2.7  & 30000 & 10 & 2305--3110 \\
\enddata
\tablenotetext{a}{All observations were taken through the
  $0\farcs2\times 0\farcs06$ apertures on 2002 August 16-17.}
\end{deluxetable}


\begin{deluxetable}{lcccccc}
\tablenum{3}
\tablecolumns{7}
\tablewidth{0pt}
\tablecaption{\HI\ Observations in the Direction of 
  vZ 1128 \label{tab:hi}}
\tablehead{
\colhead{Source} & 
\colhead{Beam} &
\colhead{$\log N(\mbox{\HI})$\tablenotemark{a}} & 
\colhead{Source\tablenotemark{b}}
}
\startdata
NRAO 21 cm    & 21\arcmin & $20.058\pm0.003$ & 1 \\
LDS 21 cm     & 35\arcmin & $20.044\pm0.005$ & 1 \\
LDS 21 cm     & 35\arcmin & $20.017\pm0.006$ & 2 \\
STIS \lyalpha &  \nodata  & $19.98\pm0.03$   & 3 \\
\enddata
\tablenotetext{a}{The uncertainties quoted for the 21 cm observations
  do not contain a contribution from errors in the stray radiation or
  baseline corrections.  These are likely of order 0.04 dex for this
  sight line.}
\tablenotetext{b}{Sources: (1) This work, through direct integration
  of the brightness temperature profile; (2) Wakker et al. 2003
  through a Gaussian decomposition of the profile; (3) This work,
  through a fit to the \lyalpha\ profile (see text).}
\end{deluxetable}


\begin{deluxetable}{lcccccc}
\tablenum{4}
\tablecolumns{7}
\tablewidth{0pt}
\tablecaption{Selected Interstellar Absorption 
  Line Measurements toward vZ 1128\tablenotemark{a}
\label{tab:measurements}}
\tablehead{
\colhead{Species} & 
\colhead{$\lambda_c$ [\AA]\tablenotemark{b}} &
\colhead{$\log \lambda f$\tablenotemark{c}} & 
\colhead{$v_-,v_+$\tablenotemark{d}} & 
\colhead{$W_\lambda$ [m\AA]} &
\colhead{$\log N_a$} &
\colhead{S/N}
}
\startdata
\ion{S}{1} & 1425.030 & 2.437 &  
     \nodata & $<15$ & $<12.7$ & 16 \\
\ion{S}{2} & 1250.584 & 0.834 & 
     $-50,35$ & $102\pm4$ & $15.25\pm0.02$ & 19 \\
\ion{S}{2} & 1253.811 & 1.135 & 
     $-75,35$ & $162\pm5$ & $15.24\pm0.02$ & 19 \\
\ion{S}{2} & 1259.519 & 1.311 & 
     $-75,35$ & $190\pm6$ & $15.20\pm0.03$ & 15 \\
\ion{Fe}{1} & 2167.453 & 2.512 &  
    \nodata & $<21$ & $<12.5$ & 18 \\
\enddata
\tablenotetext{a}{Measurements from STIS E140M and E230M observations
  of \vz.  All limits are $3\sigma$ and uncertainties include
  contributions from photon statistics and an assumed zero-point
  uncertainty of $2\%$ of the local continuum.  Signal-to-noise
  ratios, S/N, determined empirically following Sembach \& Savage
  1992.}
\tablenotetext{b}{Rest wavelength of the transition, from the
  compilation of Morton 2003.}
\tablenotetext{c}{Oscillator strength, $f$, given as $\log \lambda f$,
  from the compilation of Morton 2003.}
\tablenotetext{d}{\hspace{0.01in}LSR velocity range of integration for quoted values
  of equivalent width and apparent column density.}

\end{deluxetable}


\begin{deluxetable}{llc}
\tablenum{5}
\tablecolumns{3}
\tablewidth{0pt}
\tablecaption{Adopted Column Densities Toward M~3
\label{tab:columns}}
\tablehead{
\colhead{Species} & 
\colhead{$\log N$} &
\colhead{Ref.\tablenotemark{a}}
}
\startdata
\ion{H}{1} & $19.98\pm0.03$ & 1 \\
$e^-$      & $ 19.912\pm0.002$ & 2 \\
H$_2$      & $<14.35 \, (3\sigma)$ & 1 \\
\ion{O}{6} & $14.49\pm0.03$ & 1 \\ 
\ion{S}{1} & $<12.7 \, (3\sigma)$ & 3 \\
\ion{S}{2} & $15.28\pm0.02$ & 3 \\
\ion{S}{3} & $14.47\pm0.03$ & 1 \\
\ion{S}{4} & $<13.7 \, (3\sigma)$ & 1 \\
\ion{S}{6} & $<13.4 \, (3\sigma)$ & 1 \\
\ion{Fe}{1} & $<12.5 \, (3\sigma)$ & 3 \\
\ion{Fe}{2} & $14.80\pm0.05$ & 1 \\
\ion{Fe}{3} & $14.42\pm0.05$ & 1 \\
\enddata
\tablenotetext{a}{References: (1) Howk et al. 2003 (Paper~I); (2)
  Ransom et al. 2004; (3) this work.}
%
%
\end{deluxetable}


\begin{deluxetable}{lrr}
\tablenum{6}
\tablecolumns{3}
\tablewidth{0pt}
\tablecaption{Derived Interstellar Parameters Toward M~3
\label{tab:results}}
\tablehead{
\colhead{} & 
\multicolumn{2}{c}{Helium Ionization} \\
\cline{2-3} 
\colhead{Quantity} &
\colhead{Minimum\tablenotemark{a}} & 
\colhead{Maximum\tablenotemark{b}}
}
\startdata
%
%
$\eta\subwim\,$\tablenotemark{c} & $0.98\pm0.01$  & $0.81\pm0.04$  \\

$\log \Nelectron$        & \multicolumn{2}{c}{$19.912\pm0.002$} \\

$\log \Nelectron\subhim$ & \multicolumn{2}{c}{$19.10\pm0.12$} \\

$\log \Nelectron\subwim$ & \multicolumn{2}{c}{$19.84\pm0.03$} \\

$\log N(\mbox{\HII})\subhim$ 
                        & \multicolumn{2}{c}{$19.01\pm0.18$} \\

$\log N(\mbox{\HII})\subwim$
                        & $19.83\pm0.04$ & $19.75\pm0.04$ \\

$\log N({\rm H})$\tablenotemark{d}
                        & $20.22\pm0.02$ & $20.19\pm0.03$ \\
$\log N({\rm H})_{Total}$\tablenotemark{e}
                        & $20.24\pm0.03$ & $20.21\pm0.03$ \\
$N(\mbox{\ion{H}{2}})\subwim/N({\rm H})$\tablenotemark{d}
                        & $0.42\pm0.04$  & $0.37\pm0.04$  \\  
$N(\mbox{\ion{H}{2}})_{Total}/N({\rm H})_{Total}$\tablenotemark{e}
                        & $0.45\pm0.05$  & $0.41\pm0.06$  \\  
$\log N({\rm S})$       & \multicolumn{2}{c}{$15.34\pm0.02$} \\
$\log N({\rm Fe})$      & \multicolumn{2}{c}{$14.95\pm0.04$} \\

$\log A({\rm S})$       & $-4.87\pm0.03$ & $-4.84\pm0.03$ \\
$\log A({\rm Fe})$      & $-5.27\pm0.05$ & $-5.23\pm0.05$ \\
%
%
%
%
%
\enddata 
\tablenotetext{a}{The minimum helium ionization case assumes $x({\rm
    He}^{+2})\subwim=0$. We prefer this assumption over the maximum
  helium ionization case.}
\tablenotetext{b}{The maximum helium ionization case assumes $x({\rm
    He}^{+2})\subwim=1-x({\rm He}^{+})\subwim$.}
\tablenotetext{c}{The values of $\eta\subwim$ are derived assuming
  $0.67 \la x({\rm H}^+)\subwim \la 1.0$ and $x({\rm He}^{+})\subwim
  \la 0.27 \, x({\rm H}^+)\subwim$.  The uncertainties quoted
  encompass the range of values derived using these constraints and
  the assumptions regarding $x({\rm He}^{+2})\subwim$ appropriate for
  each model.}
\tablenotetext{d}{These quantities refer only to warm gas, including
  the WNM and WIM.  They exclude ionized gas associated with the HIM.}
\tablenotetext{e}{The quantities with the subscript ``Total'' include
  contributions from both warm and hot gas, i.e., they include ionized
  gas associated with the HIM.}
\end{deluxetable}



\end{document}